\documentclass{webofc}

\usepackage[varg]{txfonts}   
\usepackage{hyperref}
\usepackage{url}
\hypersetup{colorlinks=true,citecolor=blue,urlcolor=blue,linkcolor=blue}

\begin{document}

\title{Leading order, next-to-leading order, and non-perturbative parton collision kernels: Effects on the jet substructure}

\author{
\firstname{Rouzbeh} \lastname{Modarresi-Yazdi}
\inst{1}\fnsep\thanks{\email{rouzbeh.modarresi-yazdi@mail.mcgill.ca}}
\and
\firstname{Shuzhe} \lastname{Shi}
\inst{2,3}\fnsep\thanks{\email{shuzhe-shi@tsinghua.edu.cn}(Speaker)} 
\and
\firstname{Charles} \lastname{Gale}
\inst{1}\fnsep\thanks{\email{charles.gale@mcgill.ca}}
\and
\firstname{Sangyong} \lastname{Jeon}
\inst{1}\fnsep\thanks{\email{sangyong.jeon@mcgill.ca}}
}
\institute{
Department of Physics, McGill University, 3600 University Street, Montreal, QC, Canada H3A 2T8. 
\and
Department of Physics, Tsinghua University, Beijing 100084, China.
\and
State Key Laboratory of Low-Dimensional Quantum Physics,
Tsinghua University, Beijing 100084, China.
}

\abstract{As an important signature of the quark-gluon plasma (QGP), a high-precision energy-loss model is essential to independently verify the QGP properties extracted from soft particles. In this work, we  optimize the energy loss modeling in MARTINI by introducing the formation time of the parton shower in the initial hard scattering, which is essential for a simultaneous description of the hadron and jet $R_{AA}$. Based on this, we study the phenomenological influence of the higher order collision kernels --- the up-to-NLO one evaluated by EQCD and the non-perturbative (NP) one computed in lattice QCD --- on the energy loss of the hard parton, compared to the LO kernel. The hadron and jet $R_{AA}$ are calculated with AMY rates using the three kernels and the optimized parameter sets for the running coupling. The results exhibit remarkable similarities in their overall values as well as in $p_T$ and centrality dependence. Sizable differences in the jet substructure are observed between different soft radiation rates.
}
\maketitle
\section{Introduction}\label{sec:introduction}
The production and characterization of hot, dense strongly interacting matter is a central focus in subatomic physics, aiming to map the Quantum Chromodynamics (QCD) phase diagram. A key discovery in this field is the quark-gluon plasma (QGP), a state of matter predicted by non-perturbative QCD calculations and confirmed through experiments at the Relativistic Heavy-Ion Collider (RHIC) and the Large Hadron Collider (LHC). The study of QGP has entered a precision era, with researchers using differential observables and penetrating probes to characterize its properties. The temperature of the plasma can be inferred from electromagnetic probes \cite{Gale:2025ome}, and valuable characterization is also provided by the study of QCD jets: sprays of particles resulting from the decay of energetic partons formed early in collisions. These jets, reconstructed from correlated hadrons in the final state, provide insight into  medium properties by revealing how they interact with and lose energy.

Theoretical models have evolved to account for the rapidly changing nature of the QGP, which undergoes multiple stages of evolution, some far from equilibrium. Models like MARTINI~\cite{Schenke:2009gb}, LBT~\cite{Zhang:2003yn, He:2015pra, Cao:2016gvr}, CUJET~\cite{Buzzatti:2011vt, Shi:2018lsf}, and AdS-CFT~\cite{Casalderrey-Solana:2014bpa} simulate jet evolution against a fluid dynamical background, incorporating energy and momentum exchange between partons and the medium. These models rely on various formalisms, such as AMY~\cite{Arnold:2002ja}, higher twist~\cite{Qiu:1990xxa}, and DGLV~\cite{Gyulassy:1993hr}, to describe the energy loss of the jet. For recent reviews of the various frameworks and models of jet energy loss, see Refs.~\cite{Cao:2024pxc,gale:2013da} and references therein.

\section{Jet energy loss in MARTINI}\label{sec:eloss.theory}
The MARTINI framework~\cite{Schenke:2009gb} is a Monte Carlo solution to the rate equation for evolving hard parton distributions, describing energy gain and loss:
\begin{equation}
\begin{split}
    \frac{df}{dt}(p) = \int_{-\infty}^{\infty} dk\,
    \bigg( \frac{d\Gamma(p+k,k)}{dk} f(p+k)
  - \frac{d\Gamma(p,k)}{dk} f(p)\bigg),
\end{split}
\end{equation}
where $ f(p) $ is the parton distribution and $ d\Gamma/dk $ are energy loss rates. MARTINI includes both radiative (gluon bremsstrahlung) and elastic energy loss channels.

The primary energy loss mechanism is the radiative channel, computed using the AMY formalism~\cite{Arnold:2001ba}. The inelastic rates are given by:
\begin{equation}
    \begin{split}
        \frac{\mathrm{d} \Gamma_{i\to jk}}{\mathrm{d} x} (p,x) =\; &
        \frac{\alpha_s P_{i \to jk}(x)}{[2p\,x(1{-}x)]^2} \bar{f}_j(x\,p)\, \bar{f}_k((1-x)p)  \int \! \frac{\mathrm{d}^2 \mathbf{h}_{\perp}}{(2\pi)^2} ~\text{Re} \left[ 2\mathbf{h}_{\perp} \cdot \mathbf{g}_{(x,p)}(\mathbf{h}_{\perp}) \right].
        \label{eq:splitting.rate.martini}
    \end{split}
\end{equation}
Here, $ \bar{f}_{j} = (1\pm f_{j}) $ accounts for Bose enhancement or Pauli suppression, $ P_{i\to jk}(x) $ is the DGLAP splitting kernel, and $ \mathbf{g}_{(x,p)}(\mathbf{h}_{\perp}) $ solves:
\begin{align}
    \begin{split}
        2\mathbf{h}_{\perp} =\; & i \delta E(x,p,\mathbf{h}_{\perp}) \mathbf{g}_{(x,p)}(\mathbf{h}_{\perp})
        + \int \frac{\mathrm{d}^2\mathbf{q}_{\perp}}{(2\pi)^2}~\bar{C}(q_\perp)  \Big\{ C_{1} [ \mathbf{g}_{(x,p)}(\mathbf{h}_{\perp}) - \mathbf{g}_{(x,p)}(\mathbf{h}_{\perp} -\mathbf{q}_{\perp}) ] \nonumber \\
& + \, C_{x} [\mathbf{g}_{(x,p)}(\mathbf{h}_{\perp}) - \mathbf{g}_{(x,p)}(\mathbf{h}_{\perp} -x\mathbf{q}_{\perp}) ] 
 + \, C_{1-x} [\mathbf{g}_{(x,p)}(\mathbf{h}_{\perp}) - \mathbf{g}_{(x,p)}(\mathbf{h}_{\perp} -(1{-}x)\mathbf{q}_{\perp}) ]\Big\}.
    \end{split}
    \label{eq:AMY.linear.integral.equation}
\end{align}
The energy difference $ \delta E(x,p,\mathbf{h}_{\perp}) $ and collision kernel $ \bar{C}(q_\perp) $ are key components. The leading-order (LO) collision kernel is:
\begin{align}
    \bar{C}_{\mathrm{LO}}\left(\mathbf{q}_{\perp}\right) = \frac{g^2 T^3}{q^2_{\perp}\left(q^2_{\perp} + m^2_D\right)}\int \frac{\mathrm{d}^3p}{\left(2\pi\right)^3} \frac{p-p_z}{p} [2 C_A f_B(p)\bar{f}_B(p')+ 4N_{f} T_f f_F(p)\bar{f}_F(p')]\,,
\end{align}
where $ m_D $ is the Debye mass, and $ f_F $, $ f_B $ are Fermi--Dirac and Bose--Einstein distributions. Next-to-leading-order (NLO) and non-perturbative (NP) kernel evaluations have also been developed~\cite{Caron-Huot:2008zna, Moore:2021jwe}.

The strong coupling $ \alpha_s $ runs according to:
\begin{equation}
    \alpha_s(\mu^2) = \frac{4\pi}{\left(11 -\frac{2}{3} N_f\right)\log{\left(\frac{\mu^2}{\Lambda^2_{\mathrm{QCD}}}\right)}}\,,
    \label{eq:alphas.running.lo.expression}
\end{equation}
with $ \Lambda_{\mathrm{QCD}} = 200\mathrm{\;MeV} $. The renormalization scale $ \mu $ is set by:
\begin{equation}
    \mu = \sqrt{\langle p^2_T\rangle}=\begin{cases}
        \kappa_e \sqrt{\hat{q} \lambda_{\mathrm{mfp}}} & \mathrm{elastic\;\&\; conv. \;channels} \\
        \kappa_r \left(\hat{q}p\right)^{1/4}           & \mathrm{radiative\;channels}
    \end{cases}.
    \label{eq:alphas.scales}
\end{equation}
These parameters are central to MARTINI's energy loss modeling. Given that the $\kappa$ values multiply the renormalization scale, a larger $\kappa_{r,e}$ implies a smaller $\alpha_s$ for the respective channel. The value of those parameters is extracted from data. 

In this work, we modify the above picture of hard parton generation and evolution in MARTINI framework to also account for a ``shower formation'' time, post hard scattering event. In order to study this effect we assign a formation time to each outgoing hard parton \cite{Zhang:2022ctd} by summing over the splitting time of its parent partons,
\begin{equation}
    \tau_{\mathrm{form.}} = \sum_i \tau_{\mathrm{form.}, i}\,\qquad
    \tau_{\mathrm{form.}, i} = \frac{2E_i\;x_i(1-x_i)}{k^2_{\perp, i}},
    \label{eq:form.time}
\end{equation}
where $E_i$ is the energy of the parent, $x_i$ and $1-x_i$ the energy fractions taken by the outgoing particles and $k_{\perp, i}$ is the transverse momentum of the outgoing partons relative to the parent. Thus each parton is only allowed to interact with the local thermal medium if the current time $\tau$ is greater than both the formation time of the particle and the start time of the hydro, $\tau > \max{(\tau_0, \tau_{\mathrm{form.}})}$. 

\section{Results}
\subsection{Importance of shower formation time}
\begin{figure*}[!hbtp]\centering
\includegraphics[width=0.48\textwidth,clip]{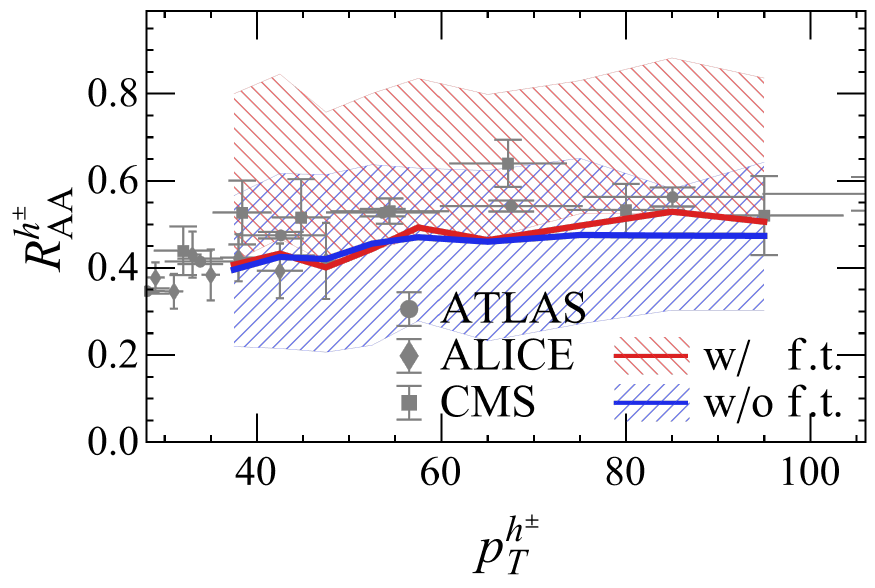}
\includegraphics[width=0.48\textwidth,clip]{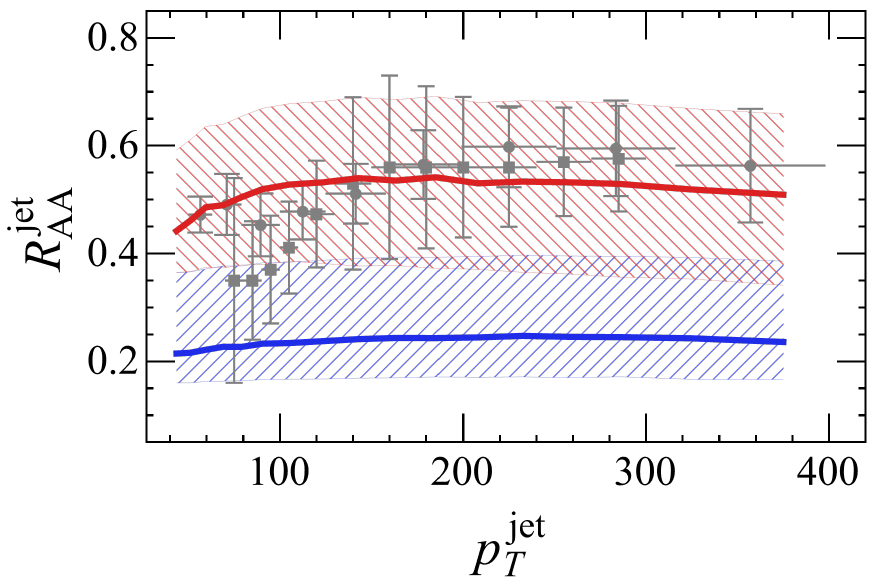}
\caption{Inclusive charged hadron (left) and jet (right) $R_{AA}$ for $0$-$5\%$ most central $\mathrm{Pb}+\mathrm{Pb}$ collisions at $\sqrt{s}=2.76~\mathrm{ATeV}$. 
Results with (without) shower formation time are colored in red (blue). The shaded areas indicate upper/lower limit covered by the parameter range $\kappa_{r,e}\in [0, 15]$, whereas the thick lines correspond to the most optimal parameter set. Data are from the ALICE~\cite{ALICE:2012aqc}, ATLAS~\cite{ATLAS:2015qmb} and CMS~\cite{CMS:2012aa, CMS:2016uxf} Collaborations.}
\label{fig-1}
\end{figure*}
We begin by showing the importance of introducing the formation-time of final state parton shower in MARTINI. We focus on the LO results for $0$-$5\%$ most central $\sqrt{s}=2.76~\mathrm{TeV}$ Pb+Pb collisions in this section, but the NLO and NP rates and calculation in other centrality bins tell the same story. As mentioned in the preceding section, the important parameters of MARTINI are those that govern the running of the strong coupling, i.e., $\kappa_r$ and $\kappa_e$, the coefficients of the renormalization scale of the strong coupling in radiative and elastic scattering channels, respectively. A group of fifty pairs of $(\kappa_r,\kappa_e)$ is randomly and evenly sampled from the interval $\kappa_{r,e}\in [0, 15]$. 

Let us first consider the nuclear modification factors for charged hadron and jets,
\begin{equation}
    R^{h^{\pm}}_{\mathrm{AA}} = \frac{dN^{h^{\pm}}_{\mathrm{AA}}/dp_Td\eta}{N_{\mathrm{bin.}}dN^{h^{\pm}}_{\mathrm{pp}}/dp_Td\eta}\;,
    \qquad
    R^{\mathrm{jet}}_{\mathrm{AA}} = \frac{dN^{\mathrm{jet}}_{\mathrm{AA}}/dp_Td\eta}{N_{\mathrm{bin.}}dN^{\mathrm{jet}}_{\mathrm{pp}}/dp_Td\eta}\,,
\end{equation}
where $N_{\mathrm{bin.}}$ is the number of binary collisions. 
Throughout this work, jets are clustered using the anti-$k_T$ algorithm~\cite{Cacciari:2008gp} for different radii of the jet cone $R = \sqrt{\Delta \eta^2 + \Delta \phi^2}$, where $\Delta \eta\equiv~\eta_i~-~\eta_j$ and $\Delta \phi\equiv \phi_i-\phi_j$ are the difference in pseudorapidity and azimuthal angle (in momentum space) between different hard partons ($i$ and $j$) during the clustering process~\cite{Cacciari:2008gp}.

The results are shown in figure~\ref{fig-1}, with the shaded areas indicating upper/lower limit covered by the parameter range $\kappa_{r,e}\in [0, 15]$. One may see that while the bands exhibits good coverages of $R^{h^{\pm}}_{\mathrm{AA}}$ in both models, the $R^{\mathrm{jet}}_{\mathrm{AA}}$ results indicate significant over-quenching when shower formation time is not turned on: none of the $\kappa$ pairs are able to match the data. The model-data comparison is immediately improved when formation time effect is induced.  The over-quenching in the former case is indicative of long evolution time. By considering a low-virtuality model and having an instantaneous parton shower, all partons coming out of the hard interaction point are confronted with an evolving medium and see the whole evolution history.
The comparison is more obvious in the most optimal parameter set represented by the thick lines, with details of parameter optimization can be found in our paper~\cite{Modarresi-Yazdi:2024vfh}.

\subsection{Comparison of LO/NLO/NP collision kernels}
\begin{figure*}[!hbtp]\centering
\includegraphics[width=0.24\textwidth,clip]{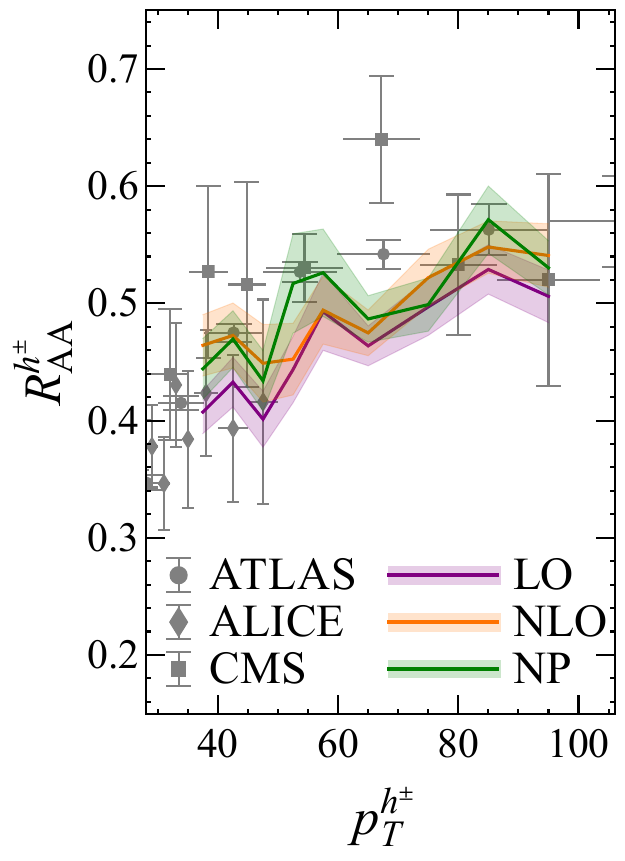}
\includegraphics[width=0.24\textwidth,clip]{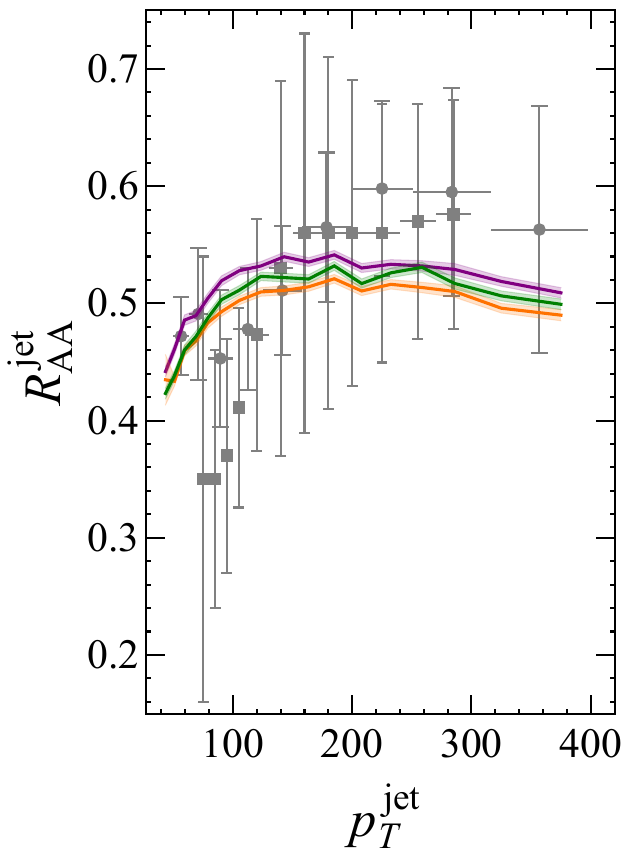}
\includegraphics[width=0.24\textwidth,clip]{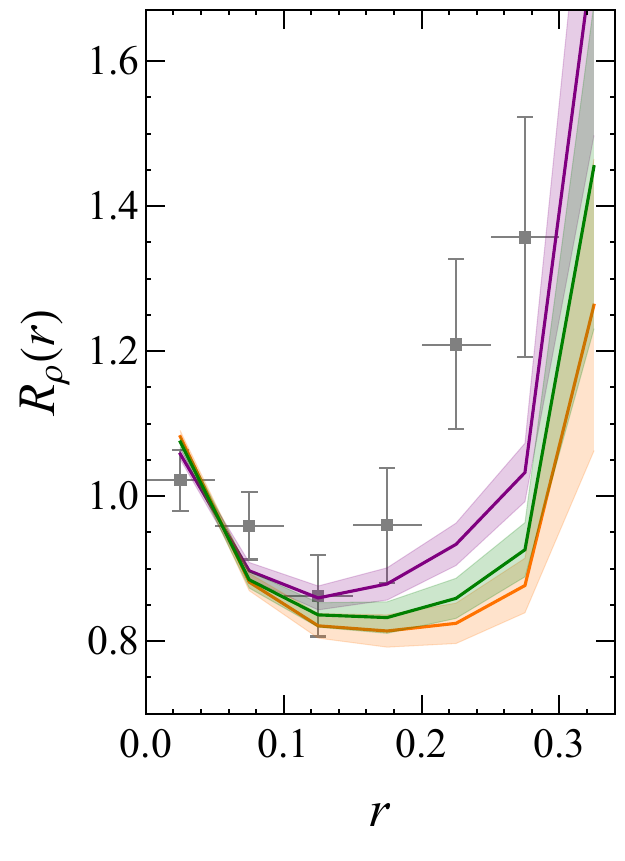}
\includegraphics[width=0.24\textwidth,clip]{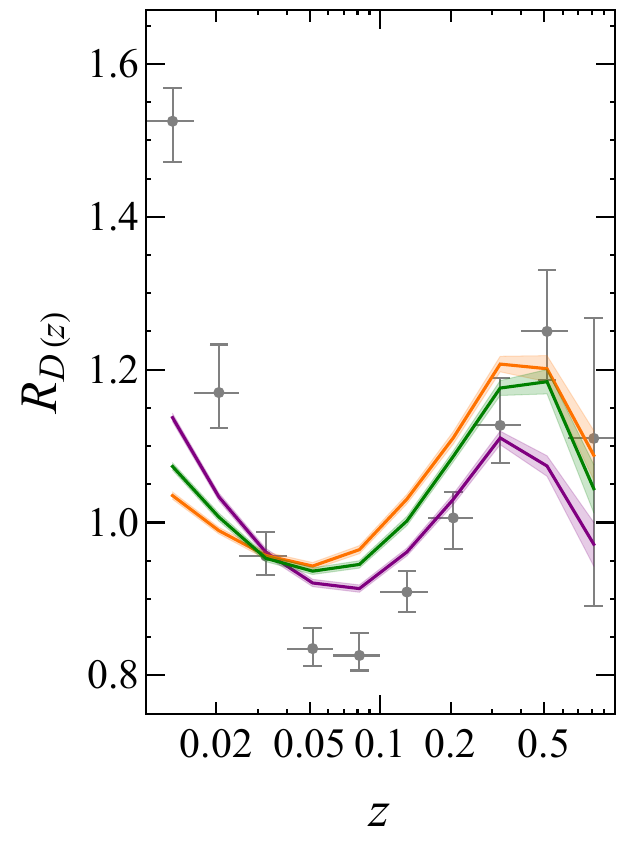}
\caption{Comparison of simulation results using LO (purple), NLO (orange), and NP (green) collisions kernels of (from left to right) inclusive charged hadron $R_{AA}$, inclusive jet $R_{AA}$, jet shape ratio, and jet fragmentation function ratio for $0$-$5\%$ most central $\mathrm{Pb}+\mathrm{Pb}$ collisions at $\sqrt{s}=2.76~\mathrm{ATeV}$. 
Formation time effect in the final state, high-virtuality parton shower is included.
Data are from the ALICE~\cite{ALICE:2012aqc}, ATLAS~\cite{ATLAS:2015qmb, ATLAS:2017nre}, and CMS~\cite{CMS:2012aa, CMS:2013lhm, CMS:2016uxf} Collaborations.}
\label{fig-2}
\end{figure*}
With formation time effect properly implemented, we separately tune the coupling scale for rate sets using different collision kernels and find $\kappa_r = 1.0$, $\kappa_e = 2.5$ for LO, $\kappa_r = 1.6$, $\kappa_e = 6.8$ for NLO, and $\kappa_r = 1.4$, $\kappa_e = 3.8$ for NP. The corresponding results are presented in figure~\ref{fig-2}. We observe that the overall nuclear modification factors for both charged hadrons and jets are insensitive to the collisions kernels, once the coupling parameters are fitted respectively. This indicates that the difference in radiation rates can be absorbed by a rescaling of $\alpha_s$, as also indicated in~\cite{Yazdi:2022bru}.

Different collisions rates are distinguishable in more differential jet substructure observables such as the AA-to-pp ratios of jet shape, the hadron distribution with respect to the transverse distance $r = ((\phi_\mathrm{trk} - \phi_\mathrm{jet})^2 + (y_\mathrm{trk}-y_\mathrm{jet})^2)^{\frac{1}{2}}$ and fragmentation function, that with respect to the longitudinal momentum fraction $z \equiv \frac{\mathbf{p}_\mathrm{jet} \cdot \mathbf{p}_\mathrm{trk}}{\mathbf{p}_\mathrm{jet} \cdot \mathbf{p}_\mathrm{jet}}$,
\begin{align}
R_{\rho} = & 
    \frac{\rho_{\mathrm{AA}}(r)}{\rho_{\mathrm{pp}}(r)}\,,\qquad
    \rho(r) \equiv \frac{N_\mathrm{norm}}{N_\mathrm{jet}} \frac{\sum_\mathrm{jets} \sum_{r \in [r_\mathrm{min},r_\mathrm{max})} {p_{T}^{\mathrm{trk}}}/{p_{T}^{\mathrm{jet}}}}{r_\mathrm{max} - r_\mathrm{min}}\,,\\
R_{D(z)} = &
    \frac{D(z)_{\mathrm{AA}}}{D(z)_{\mathrm{pp}}}\,,\qquad
    D(z)
        \equiv
        \frac{\sum_\mathrm{jets} \sum_{z \in [z_\mathrm{min},z_\mathrm{max})}1}{N_\mathrm{jet} \; (z_\mathrm{max} - z_\mathrm{min})}
        \,.
\end{align}
We observe a stronger sensitivity to the models. The three models show excellent agreement near the jet axis ($r \leq 0.1$), but a clear separation emerges in the outer regions, with the higher-order kernel models (NLO and NP) diverging from the LO model.

The differences in the rates manifest themselves in how they populate lower-energy modes or their tendency to lose energy through soft radiation relative to the jet energy. Specifically, the NLO and NP rates produce more soft partons, which are pushed outside the jet cone, whereas the LO model does not exhibit this behavior. This aligns with the findings in Ref.~\cite{Yazdi:2022bru} for a fixed-$\alpha_s$ scenario. In that study, after rescaling by a constant factor, the NLO and NP rates could approximate the LO rate for hard radiation, but significant discrepancies persisted for soft radiation.

\section{Summary}
Based on published results~\cite{Yazdi:2022bru, Modarresi-Yazdi:2024vfh}, this proceeding reports quantitative model comparison of two technical aspects of jet evolution in relativistic heavy-ion collisions. First, it examines the early-time dynamics of jet formation, distinguishing between two scenarios: (1) a "single-stage" model, where the parton shower fully develops before hydrodynamic evolution begins, and (2) a "time-delayed" model, where the shower evolves over time, with some partons shedding virtuality during or after QGP formation. The latter is crucial for interpreting jet-related observables. Second, the study investigates the impact of up-to NLO and NP collision kernels on AMY radiative rates. This refinement aims to improve the accuracy of jet energy loss simulations in both static and dynamically evolving QGP environments.

\vspace{3mm}
\textbf{Acknowledgments} --- The authors acknowledge useful discussions with Shanshan Cao, Matthew Heffernan, Weiyao Ke, Amit Kumar, and Abhijit Majumder. This work is supported in part by the Natural Sciences and Engineering Research Council of Canada (RMY, CG, SJ), in part by Tsinghua University under grant No. 53330500923 (SS). Computations were carried out on the Narval, Beluga and Graham clusters managed by Calcul Qu\'ebec and the Digital Research Alliance of Canada.

\bibliography{ref}

\end{document}